\documentclass[iop]{emulateapj}
\usepackage{natbib}
\usepackage{amsmath}
\newcommand\mA{$\rm{m\AA}$ }

\def\ls{{_<\atop^{\sim}}}

\newcommand\kms{\rm{km\,s^{-1}}}

\begin{document}

\title{An X-ray WHIM metal absorber from a Mpc-scale empty region of space}
\shorttitle{WHIM at z=0.11}
\shortauthors{Zappacosta et al.}
\author{L. Zappacosta\altaffilmark{1}, F. Nicastro\altaffilmark{1,2}, Y. Krongold\altaffilmark{3}, R. Maiolino\altaffilmark{4}}
\affil{INAF - Osservatorio Astronomico di Roma, via di Frascati 33, 00040 Monte Porzio Catone, Italy}
\altaffiltext{1}{INAF - Osservatorio Astronomico di Roma, via di Frascati 33, 00040 Monte Porzio Catone, Italy; luca.zappacosta@oa-roma.inaf.it}
\altaffiltext{2}{Harvard-Smithsonian Center for Astrophysics, 60
  Garden Street, Cambridge, MA 02138}
\altaffiltext{3}{Instituto de Astronomia, Universidad Nacional
  Autonoma de Mexico, Apartado Postal 70-264, 04510 Mexico DF, Mexico}
\altaffiltext{4}{Astrophysics Group, Cavendish Laboratory,
  19 J.J. Thomson Avenue, Cambridge, CB3 OHE}
\begin{abstract}
We report a detection of an absorption line at $\sim44.8~\rm\AA$ in a $>500~\rm ks$ Chandra HRC-S/LETG 
X-ray grating spectrum of the blazar H~2356-309. This line can be identified
as intervening $\rm C\,V$-K$\alpha$ absorption, at $\rm z\approx0.112$, produced by a warm 
($\log \rm T = 5.1~\rm K$) intergalactic absorber. The feature is
significant at a $2.9\sigma$ level (accounting for the
number of independent redshift trials). 
We estimate an equivalent hydrogen column density of 
$\log \rm N_H=19.05~(Z/Z_\odot)^{-1}~\rm cm^{-2}$. 
Unlike other previously reported FUV/X-ray metal detections of warm-hot intergalactic medium (WHIM), this CV 
absorber lies in a region with locally low galaxy density, at $\sim2.2~\rm Mpc$ from the closest galaxy at that redshift, and therefore 
is unlikely to be associated with an extended galactic halo. 
We instead tentatively identify this absorber with  an intervening Warm-Hot 
Intergalactic Medium filament  possibly permeating a large--scale, 30~Mpc 
extended, structure of galaxies whose redshift centroid, within a cylinder of 7.5 Mpc radius centered on the 
line of sight to H 2356-309, is marginally consistent (at a $1.8~\sigma$ level) with the redshift of the absorber.
\end{abstract}


\section{Introduction}
Cosmological simulations predict that during the process of structure formation, the high redshift diffuse cold photoionized 
intergalactic baryonic phase responsible for the $Ly\alpha$ forest, undergoes a substantial transformation. 
Baryons in these filaments are predicted to fuel structure formation by accreting toward the large--scale dark matter Cosmic Web, populated by galaxies and, by doing so, to undergo shocks that heat them up 
to temperatures of $\rm{T}=10^{5-7}$~K\citep{cen,dave}. 
This new collisionally ionized baryonic phase is dubbed Warm-Hot Intergalactic Medium (WHIM), and is supposed to exist 
in filamentary structures at very low densities ($5-100$ times the mean baryonic density of the universe) and to account for the majority 
of the baryons in the local universe ($z<1$) that are currently still undetected \citep[][]{fukugita}.

The range of density and temperature of the WHIM are theoretically well established within the context of a $\Lambda$-CDM Universe. 
On the contrary, the effects of AGN and galaxy wind feedback, as well as that of galaxy mergers, on WHIM filaments, 
both in the immediate surrounding of structures and at larger scales, are not yet well understood and therefore their physics 
difficult to implement in simulations. This feedback may significantly affect the metal content of the WHIM \citep[e.g.][]{cen06} and 
therefore its detectability. 
At WHIM temperatures and densities, the gas is so highly ionized and
tenuous that can only weakly emit in the soft X-rays through Breemstrahlung and weakly interact with radiation in the Far-Ultraviolet (FUV) and soft X-ray bands through electronic transitions of relatively light metals in their highest ionization states, therefore emitting or absorbing photons at these wavelenghts. 
Uncertainties on the IGM-galaxy feedback processes reflect negatively into the accuracy of hydrodynamic simulations predictions and so, in turn, on feasibility studies of WHIM observations and the prospects for accurate studies of this important baryon phase and its interplay with galaxies at large-scales.

Despite the few claims of broad-band X-ray imaging detection \citep{zappacosta,zappacosta2,werner}, the best way to currently detect and study the WHIM still relies on the identification of absorption lines (mainly OVI, OVII, CV, HI-Ly$\alpha$) from intervening medium in the FUV/X-ray spectrum of bright background extragalactic sources. 

Cross-correlation studies of WHIM metal and HI systems over the entire WHIM temperature/density distribution intervals, and their 
surrounding galaxy environments, are critical to progress in this field. 
These, however, are hampered mostly by the small number of WHIM metal detections, particularly in the regime of temperatures 
T$\gtrsim 10^{5.5}$~K (where the majority of these baryons should be found), accessible only in the X-ray band where so far only few mostly controversial results have been obtained even when targetting sightlines crossing known galaxy-traced superstructures \citep[][]{nicastro05,fang10,zappacosta10}. 
At lower temperatures (T$\simeq 10^5-10^{5.6}$~K), the WHIM can be also identified in the FUV band, through the detection 
of, possibly paired, OVI doublet and thermally broadened HI absorbers (BLAs) \citep[see][for a review]{richter}. These are found to 
constitute $\lesssim 25$\% of the predicted baryon mass, in the local Universe \citep[e.g.][]{shull11}, and 
are supposed to be tracers of collisionally ionized IGM (the WHIM), at T$\ls 10^{5.6}$ K. 
Cross-correlation studies of this cool tail of the WHIM-mass distribution with the surrounding galaxy environment, tentatively suggest 
that while OVI absorbers (or OVI and BLA pairs) lie relatively close ($\leq800~h_{70}^{-1}~\rm kpc$) to bright $L^*$ 
galaxies \citep{stocke}, BLA-only absorbers 
are found at larger distances from their nearest $L^*$ galaxy neighbor \citep[$0.75-2.9~h_{70}^{-1}~\rm Mpc$;][hereafter DSS10]{danforth}. 
This tentative evidence (based on a small number of systems) would suggest that 'warm' metals (T$\ls 10^{5.5}$) are limited to the surrounding of large galaxy halos which probably contributed to their enrichment, while the more diffuse ``metal-poor'' gas (at least in the T$\simeq 10^{5}-10^{5.6}$ temperature range) traces more distant regions.

Here, we present the first evidence for possible CV ($\lambda 40.27~\rm{\AA}$) IGM absorption in the X-rays, at z=0.112, found to lie in a relatively 
under-dense region of the local Universe, with the nearest galaxy at $\sim 2.2$ Mpc distance, but 
still within a larger-scale filamentary structure of galaxies. 
The paper is organized as follows: we briefly present the data and their reduction in \S~\ref{reduction}, then we describe the spectral 
analysis in \S~\ref{analysis} and finally in \S~\ref{discussion} we discuss and iterpret our findings. 

\noindent Reported errors are $1\sigma$ unless otherwise specified.

\section{Data reduction}\label{reduction}
H2356-309 has been observed twice with Chandra with the HRC-S/LETG grating configuration for a total of $\sim 600$~ks. 
The first observation was performed in October 2008 (96.49~ks). The second observation (496.4~ks) consists in 10 pointings 
with exposures ranging from 15~ks to 100~ks and was carried out during the last four months of
2009. 
H~2356-309 was also observed with XMM-{\em Newton} for 130 ks. However, the XMM-{\em Newton} gratings (RGSs) do not 
cover the $\lambda \gtrsim 38~\rm{\AA}$ spectral band, which we focus on in this paper. We therefore do not use the XMM-{\em Newton} 
data of H~2356-306 in this paper. 
The Chandra data have been previously analyzed and presented by \citet{fang10} and
\citet{zappacosta10}. In this paper we make use of the data reduced by \citet{zappacosta10} (see their \S 3.1, for details on the 
data reduction). 
In order to maximize the signal-to-noise ratio (S/N) of the single {\em Chandra} observations, for each observation we co-add the 
HRC-LETG positive and negative order source and background spectra.  Given the non-linearity of the HRC-LETG effective dispersion 
relation 
\footnote{See http://cxc.harvard.edu/cal/Letg/Hrc\_disp/degap.html, \\
http://asc.harvard.edu/proposer/POG/html/chap9.html\#tth\_sEc9.3.2}
, this procedure may introduce additional uncertainties ($\sigma_{cal}$) in the wavelength calibration scale 
(currently calibrated to $0.01~\rm{m\AA}$). 
Since the corrections leading to these calibrations are obtained at
the rest-frame position of strong X-ray metal electronic transitions
and the interpolation to red-shifted transitions may not be strictly
valid, we assumed throughout the spectral range a conservative
increased uncertainty of $\sigma_{cal}=0.02\,\rm\AA$. This uncertainty
will in turn increase to $\rm\sigma_{cal}^{co-add}\approx0.04\,\rm\AA$ when we
will co-add the spectra of all the observations to maximize the
S/N. In this case the final uncertainty is obtained by propagating
the single spectra uncertainties but weighting each term by the
relative contribution in counts of each observation in the 0.5-2~keV
band. In the following we will always quote the statistical errors
only, unless differently specified.

\section{Spectral analysis}\label{analysis}
For the spectral analysis we use the fitting package Sherpa \citep{sherpa}, in CIAO 4.2. 
We perform all our fits using the data weighted $\chi^2$ with the Gehrels variance function and
adopting a Nelder-Mead simplex optimization method. 

The analysis is mainly focussed on the $44-48~\rm\AA$ spectral region, not considered in \citet{fang10} and 
\citet{zappacosta10}.
In our spectral analysis and fitting procedure, we use two different approaches to the data: (a) in order to maximize the 
S/N per resolution element, we first co-add all the 11 HRC-LETG spectra together \citep[see ][for details]{zappacosta10}, 
and analyze the total 496.4~ks spectrum; (b) then, to check the reliability of our findings against possible systematics 
introduced by co-adding the single spectra, we also repeat the fitting procedures simultaneously to 6 HRC-LETG 
spectra, extracted from the 11 observations in such a way that the resulting 6 spectra have all homogeneous S/N per 
resolution element: the exposures of these 6 spectra vary between 90-110 ks each.  

\noindent 
All spectra are binned to half the HRC-S/LETG FWHM (0.025~$\rm\AA$). 
 
\subsection{Single line analysis}\label{singleanalysis}
\begin{figure}[t!]
   \begin{center}
\includegraphics{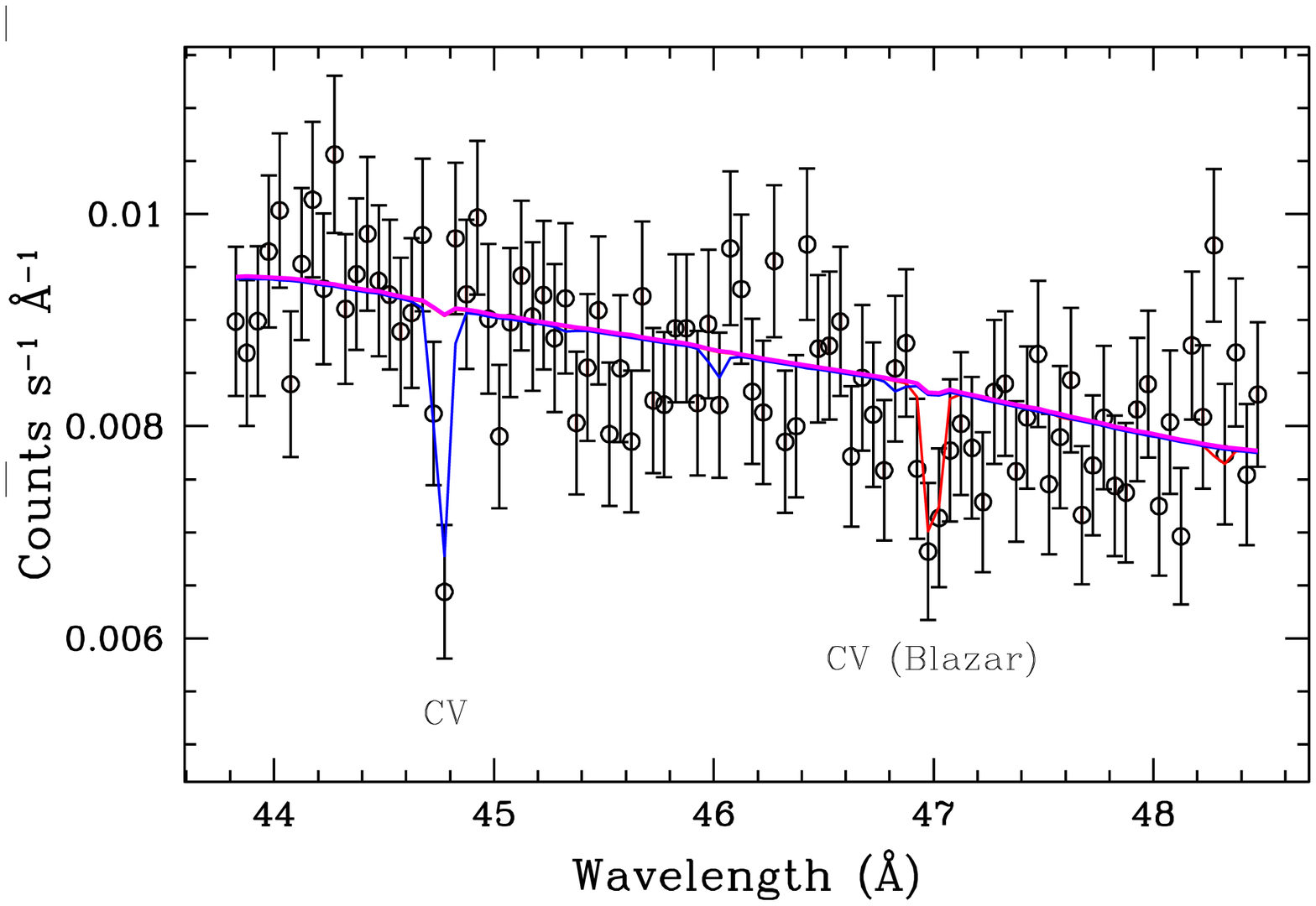}
\includegraphics{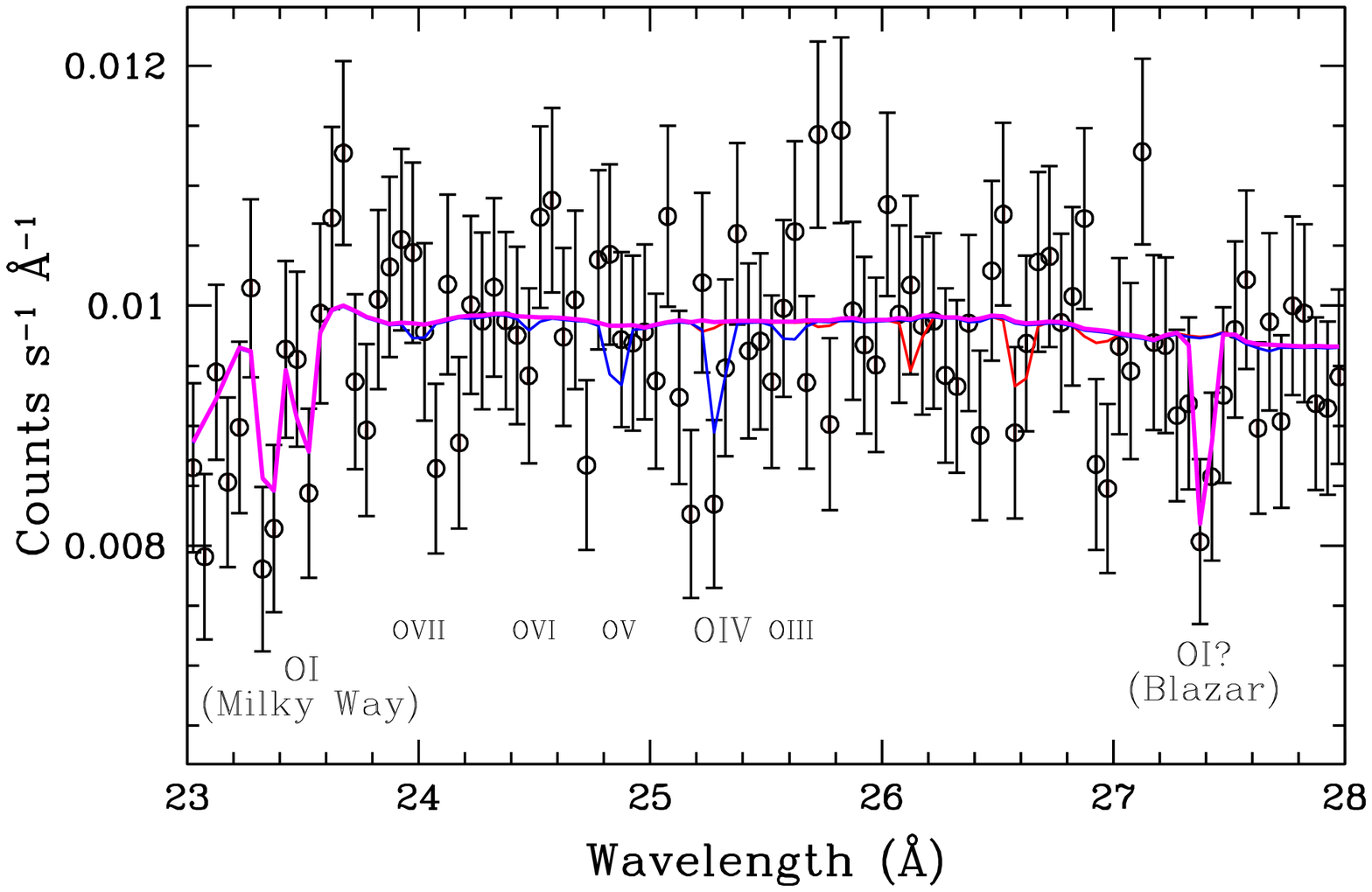}
   \end{center}
\vspace{10.5cm}
\caption{Portions of the total co-added spectrum covering the  CV and OVII regions at the z=0.112 of the intervening absorber. The magenta model is the best-fit continuum. Blue and red lines refer  to the modelling for the intervening and blazar frame systems.}
   \label{selfconsistent}
\end{figure}

Figure~\ref{selfconsistent} shows two portions of the total HRC-LETG spectrum of H~2356-309, where the redshifted CV K$\alpha$ (top panel) and 
OVII K$\alpha$ (bottom panel) lines are expected ($\lambda = 44-48.5$~$\rm\AA$, top panel, and $\lambda = 23-28$~$\rm\AA$, 
bottom panel), together with the local best fitting continuum models (magenta curves in both panels\footnote{the $\lambda = 23-28$ \AA\ best fitting continuum includes also 2 negative Gaussians to model atomic OI absorption 
both at $z=0$ and at the redshift of H~2356-309.}), and the intervening absorber models (red and blue curves, \S 3.2), folded through the HRC-LETG response. 
In these two spectral intervals, the total spectrum has $\sim 230-280$ net source counts per resolution element (CPREs; 
$\rm S/N\sim12$) and so is sensitive to the detection of absorption 
line equivalent widths $\rm{EW}\gtrsim 12\rm$~\mA at $\gtrsim 3\sigma$. 

A visual inspection of the $\lambda = 44-48.5$ \AA\ portion of the spectrum (Fig. 1, top panel) reveals the presence of two negative 
(compared to the best fitting continuum) line-like features at $\sim$44.75~$\rm\AA$ and $\sim$47~$\rm\AA$. 
We modeled these features by adding two negative Gaussians to our best fitting local continuum model. 
The two resulting best-fitting absorption lines have line centroids $\lambda = 44.76 \pm 0.01$ \AA\ and $\lambda = 47.00 \pm 0.02$ 
\AA, EW$=22\pm5$~\mA and EW$=16\pm6$~\mA, and single line statistical significances (i.e. estimated by dividing the 
best fitting Gaussian normalizations by their $1\sigma$ errors) of $4.2\sigma$ and $2.7 \sigma$, respectively. 

To check the actual statistical significance of these features against possible systematics introduced by co-adding the 11 observations, 
we performed the same analysis simultaneously on the $\lambda = 44-48.5$ \AA\ portion of the 6 homegenuous S/N spectra 
extracted from the 11 observations\footnote{ObsIDs 8120, 10764, 10497+10499, 10498+10500, 10577+10840, 10761+10762+10841}. 
For each of the 6 spectra, our local continuum model includes a
power-law with spectral index and normalizations free to vary
independently. 
We then include 2 negative Gaussians for each of the 6 spectra, with common line centroids, common EWs and 
line widths frozen to a common unresolved HRC-LETG value, and fit the 6 data-sets simultaneously. 
The best fit model gives lines positions and equivalent widths consistent with the ones derived from the fit to the co-added spectrum, 
although with larger uncertainties due to the larger number of free
parameters in the joint fit.

\begin{figure}[t!]
   \begin{center}
\includegraphics{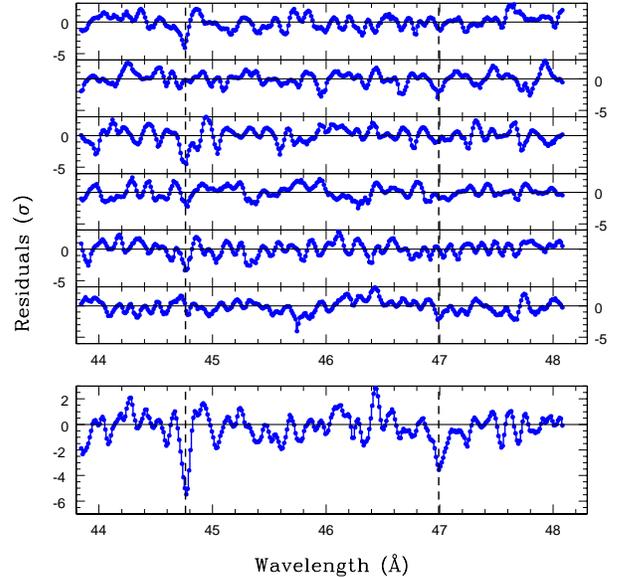}
   \end{center}
\vspace{7.9cm}
\caption{Residuals in the 44-48~$\rm\AA$ spectral region over the
  best-fit continuum model smoothed with the LETG line spread function
  and renormalized in $\sigma$ relative to their distribution. The top six panels show the 90-100~ks spectra while the bottom panel reports the total co-added spectrum.}
   \label{smoothing}
\end{figure}

Figure~\ref{smoothing} shows the smoothed residuals of the $44-48.5$ \AA\ portion of the 6 spectra (top panel) and the co-added spectrum (bottom panel), to their respective best-fitting continuum models (obtained by folding the residual histograms through the HRC-LETG line spread function), 
re-normalized after smoothing to comply with Poisson statistics. 
The $\lambda = 44.76 \pm 0.02$ \AA\ line is visible in 4 out of the 6 best-fitting continuum residuals, and appears prominent in 
the best-fitting continuum residuals to the co-added spectrum. 
The second absorption feature is only visible in the best-fitting continuum residuals to the co-added spectrum. 

In the following analysis, we focus exclusively on the co-added spectrum of H~2356-309, and tentatively identify the two absorption 
features at $\lambda = 44.76 \pm 0.02$ \AA\ and $\lambda = 47.00 \pm 0.02$ \AA, as redshifted CV K$\alpha$ lines at 
$z=0.1117 \pm 0.0005$ and $0.1671 \pm 0.0005$ (only  marginally consistent at a $3\sigma$ level with the systemic redshift of 
H~2356-309: $z=0.1654 \pm 0.0002$; \citealt{jones}).  

\subsection{Global analysis and characterization of the absorber}\label{globalanalysis}
To test the validity of our tentative CV identifications at $z=0.1117 \pm 0.0005$ and $0.1671 \pm 0.0005$, we used our 
hybrid-ionization WHIM spectral model \citep[adapted from PHASE,][]{krongold}, to check for compatibility between the data 
and the presence or absence of associated absorption from other ions of the same or different metals. 
The model includes more than 3000 electronic transitions from H to Ca, and predicts the opacity of each transition for given H 
equivalent column densities ($\rm N_H$), and temperatures (T), for collisionally ionized plasma undergoing the 
additional photoionizing contribution by UV and X-ray metagalactic photons at a given redshift and baryon density ($n_b$). 

We proceeded as follow: we first modeled local continua separately in 4 different key spectral bands of the co-added HRC-LETG 
spectrum of H~2356-309: $23.0-28.0~\rm\AA$, $31.0-36.0~\rm\AA$, $36.0-41.0~\rm\AA$, $44.0-48.5~\rm\AA$. We chose 
these 4 spectral bands because these are the regions where transitions from other abundant ions associated with CV in plasma in 
collisional ionization, photoionization, or hybrid-ionization equilibrium are expected to produce significant opacity. 
To model the local continua we used simple power-laws and, in the 23-28 \AA\ spectral range, we also included two negative 
Gaussians to model the atomic OI transitions at $z=0$ and at blazar redshift (Figure~\ref{selfconsistent}, bottom panel, magenta curves). 
Then we added two hybrid-ionization components, at the redshift of our tentative CV identifications. The model paramaters are: 
temperature, equivalent H column density and volume baryon density of the gas, the redshift of the absorber, and its internal 
turbulence velocity. The baryon volume density is used to define the relative contribution of photoionization: lower the 
volume density, higher the photo-ionization contribution from the
surrounding meta-galactic photons. This contribution 
is highly degenerate with the temperature of the plasma, and cannot be constrained independently in such quality data. 
We therefore fixed this parameter to a typical WHIM value of $\rm n_b=10^{-5} \rm cm^{-3}$ (overdensity of $\sim 50$, compared 
to the average Universe density), for both absorbers. We also froze the turbulent velocity to $\rm{v_{turb}=100~\rm\kms}$ (that of 
an unresolved line in HRC-LETG data). Of the remaining three parameters, the temperature and equivalent H column density of 
the absorbers, were left free to vary in the fit, but linked to a common value in all the 4 spectral regions.
Finally, the redshifts of the two absorbers were set to the best
fit values determined in previous section and left free to vary
within $\pm\rm{\sigma^{co-add}_{cal}}$ and linked to a common narrow spectral interval in each of the
four bands.
The best-fitting absorber models, confirm our tentative identifications of the two lines as CV K$\alpha$ absorbers. 
Fig.~\ref{selfconsistent} shows the best-fitting absorption models (blue and red curves) in the only two bands where 
relatively strong opacities from CV K$\alpha$ (top panel) and OIV-VII K$\alpha$ (bottom panel) lines are predicted. 
For both best-fitting absorber models the, by far, strongest opacity contribution is given by CV K$\alpha$. The additional opacity 
produced by OIV-VI K$\alpha$ lines is fully consistent with the data
(bottom panel, blue and red curves). 
The best-fitting parameters of the possible intervening absorbers are: $\rm z=0.1117\pm0.0005$, $\log{T}=5.10_{-0.17}^{+0.13}$~K,  
$\log{N_H}=19.05^{+0.15}_{-0.25}~(Z/Z_\odot)^{-1}~\rm{cm^{-2}}$ and $\rm z=0.1671\pm0.0005$, $\log{T}\ge4.6$~K,  
$\log{N_H}\ge18.5~(Z/Z_\odot)^{-1}~\rm{cm^{-2}}$ (90\% upper limits). 
For the intervening absorber the small upper and lower error-bars in temperature are set by the
lack of strong OVII K$\alpha$ and OIV K$\alpha$ absorption in the
data, respectively.

\noindent
We also checked whether the detected absorption feature at 44.76 \AA, could be 
due to a different transition from a cooler ion of C, at redshift lower than $z=0.112$. 
Possible candidates are inner-shell K$\alpha$ transitions from lithium-like C (CIV) all the way down 
to neutral C (CI). However, unlike CV which is Helium-like and so highly stable and practically the only ion of C over a wide range 
of temperatures (from log$T=4.8-6$ K), lower ionization ions are much less stable, and contribute together to the 
total C fraction, in a mixture that depends upon the exact gas temperature. Typically, at temperatures of log$T\sim 3-4$ K, 
CIII and CIV contribute most, while at lower temperature CI and CII are the dominant ions. 
In other words, while He-like metals can be found isolated in absorption or emission spectra, for a wide interval of temperatures, 
lower ionization ions produce equally strong opacities in groups of 2-3 ions. 
We checked the data for their compatibility with the presence of any of these low-ionization transitions, and found no possible 
solution. If the detected absorber at $\lambda=44.76$ \AA, was due, e.g. to the CIV K$\alpha$ transition at $\lambda = 41.39$ \AA\ 
(which would place the absorber at $\rm z\sim0.081$), 
in order to reproduce the spectral features at $\lambda = 44.76$ \AA, the absorber should have either a 
temperature $\log T<4$~K and $\log{N_H}\sim 19~(Z/Z_\odot)^{-1}~\rm{cm^{-2}}$ or higher temperatures but extremely high 
column densities $\log{N_H}\gtrsim20~(Z/Z_\odot)^{-1}~\rm{cm^{-2}}$. In both cases the models predicts similar, or even  higher, opacities 
at other transitions (i.e. CIII at $\sim45.55~\rm\AA$ and OIV, OVI in the range $24-25~\rm\AA$) whose presence is not consistent with 
the data. 
Similar considerations hold for the possible identification of the $\lambda = 44.76$ \AA\ feature with the main transitions 
from lower ionization ions. 
We therefore exclude the identification of the $\lambda = 44.76$ \AA\ feature with intervening cool medium at $z<0.112$, 
and conclude that a CV K$\alpha$ absorber at $z=0.1117\pm0.0005$ is the most likely identification 
for the observed feature. 

Our best-fitting model for the $z=0.1117\pm0.0005$ absorber, predicts an OVI column of 
$\sim1.3\times10^{14}~\rm{cm^{-2}}$. OVI 
has its strongest outer-shell doublet transitions in the FUV, at $\lambda\lambda1031.93,1037.62~\rm\AA$. 
We checked for the presence of these lines in the low S/N ($\sim11.5~\rm{ks}$) archival FUSE observation of H~2356-309. 
Unfortunately, however, in the relevant spectral region ($\lambda = 1147-1154$ \AA) these data are only sensitive to 
column densities of $> 1-2\times10^{14}~\rm{cm^{-2}}$ (for line Doppler parameters in the range $b=50-100$ km s$^{-1}$) at a 
$>3\sigma$ level, so fully consistent with the presence of the predicted line, but not of sufficient quality to uncontroversially set 
the issue. 

\section{Discussion}\label{discussion}

\subsection{On the significance of the intervening CV line}
In Section~\ref{singleanalysis} we estimated the significance of the intervening CV K$\alpha$ line, as a single-line 
significance of $4.2\sigma$. However, in doing so we did not account for the number of redshift trials, i.e. for the lack 
of a prior on the expected redshift of the line \citep[see e.g.][]{kaastra06}. 
A blind search for intervening CV K$\alpha$ line, exploits the entire redshift range available, from $z=0$ to the 
redshift of the background blazar. The number of resolution elements of the spectrometers between the rest frame 
position of the CV K$\alpha$ line ($\lambda_{CV}$), and the position of the line at the redshift $z_{bl}$ of the blazar, represents the number of 
independent redshift trials N. In our case: $\rm{N} =\lambda_{CV}*z_{bl}$ / 0.05 = 133. 
The probability of chance detection of an intervening $z\leq z_{bl}$ CV K$\alpha$ line seen at a single-line significance of 
$4.2\sigma$, is given by the binomial distribution formula
$$
P= \frac{N!}{n!(N-n)!} p^n (1-p)^{(N-n)}, 
$$
where $N=133$, $n=1$ and $p=1.34\times 10^{-5}$  is the
one-sided Gaussian probability corresponding to $4.2\sigma$. 
This gives a chance detection probability of $P=1.8\times 10^{-3}$
(0.2\%), corresponding to an effective line significance of 
$2.9\sigma$, after accounting for the number of redshift trials. 

\subsection{Possible galactic/super-galactic identification of the absorber} \label{ident_gals}
The low temperature estimated for the intervening CV absorber at $z=0.1117\pm 0.0005$, suggests it to be the X-ray counterpart 
of the OVI inter-galactic absorbers commonly found in the FUV spectra of extragalactic  sources, at $z\ls 0.5$ \citep[][]{danforth05,tripp08}. 
However, unlike OVI, which can be efficently produced in both photo-ionized and collisionally ionized gas, CV is an 
He-like ion, and as such much more stable and abundant in collisionally ionized gas (or WHIM) than in gas purely photo-ionized 
by the meta-galactic radiation field. 

Indeed, only a relatively small fraction of the OVI absorbers are
found to originate in shock-heated (e.g. collisionally ionized) gas at
low-temperature end of
the WHIM mass distribution and therefore associated to BLAs, while
the rest is possibly imprinted by the residual local photo-ionized
Ly$\alpha$ forest \citep[e.g.][]{danforth05,tripp08}.  The CV absorber
that we find here, belongs probably to the class of OVI-BLA
associations, which are still limited in number.

In a recent work on BLA detections along the line of sight to 7 AGNs, DSS10 report a marginal evidence for BLA-OVI detections in 
well-surveyed galaxy fields (4 in their sample) to be closer to $L^*$
galaxies ($\ls 0.5 Mpc$ from their closest galaxy) than isolated BLAs  
with OVI non-detections (4 in their sample; $0.75$ Mpc $< d < 2.9$ Mpc from the closest galaxy). 
This limited number statistic evidence, suggests that metals in the local
Universe  (at least at such moderately low temperatures) are not uniformly spread 
in the IGM, but they concentrate around structures, possibly in
extended halos of galaxies.
 
However, these searches are limited to nearest-galaxy versus absorber correlations and do not investigate on the spatial distributions 
of galaxies around a given absorbers. WHIM filaments embedding a large number of galaxies could be enriched differently, and at 
different levels, from extended galaxy halos in relatively sub-dense galaxy regions. 

\begin{figure}[!t]
   \begin{center}
     \includegraphics[scale=0.48]{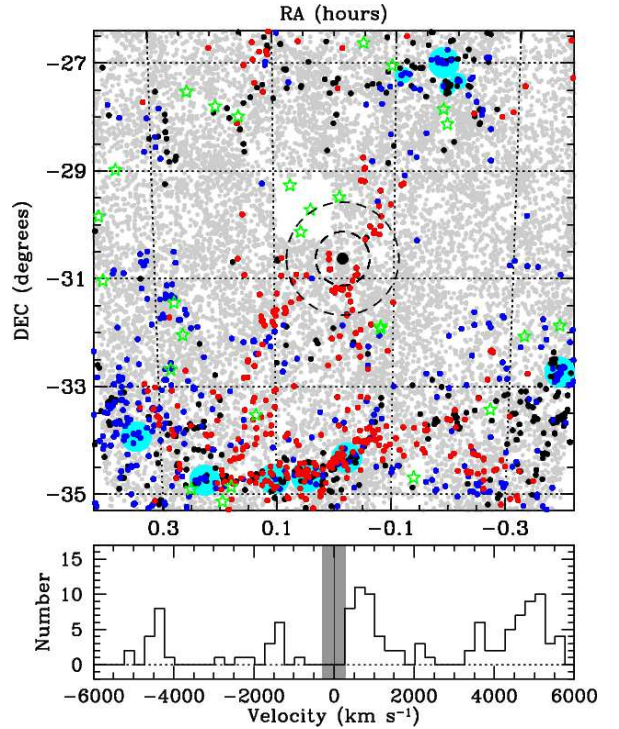}
   \end{center}
   \caption{Top panel: sky map centered at the blazar location (big black dot) showing the galaxy distribution (dots) from the 2dFGRS.  
     All the black, red and blue points show the whole galaxy distribution within $\pm1000~\rm
     \kms$ (i.e. $\sim30~\rm{Mpc}$ comoving, neglecting the galaxies
peculiar motions) from the absorber redshift. Each color represents a
redshift range, i.e. blue for $\Delta z=0.1083-0.1105$; black for
$\Delta z=0.1105-0.1127$; red for 
     $\Delta z=0.1127-0.1150$. Grey dots are galaxies at all other redshifts. The green stars mark the position of bright ($\rm{m_v<7}$) 
     foreground stars in our Galaxy. Clusters and groups of galaxies are shown as solid cyan circles with radii  
     of 2~Mpc and 1~Mpc. The two dashed circles around the position of H~2356-309, comprise two regions of $\sim 4$~Mpc and $\sim 7.5$
     Mpc radius, at the blazar's redshift. The bottom panel shows the
     velocity distribution of the galaxies relative to the CV absorber velocity in a 
     cylinder of 7.5~Mpc radius centered on the line of sight to H~2356-309. 
     The uncertainty  on the position of the CV~K$\alpha$ absorber
     (see \S\ref{ident_gals}) is shown by
     the grey area.}
   \label{pieplot}
\end{figure}

Much sparser are the evidences regarding metals at higher temperatures
that can be traced in the X-rays. 
\citet{nicastro05} and \citet{fang10} have obtained so far the most 
convincing, although still debated \citep[e.g.][respectively]{kaastra06,yao}
detections from this hotter phase. They have found absorptions from warm-hot material 
at $z=0.027$ and
$z=0.03$, both coincident with extended filaments of galaxies
\citep{williams10,buote09}, in regions not completely devoid
of $0.4-1~L^*$ galaxies at scales $\lesssim400~\rm{kpc}$
\footnote{Indeed for the z=0.027
system there are two galaxies (of 0.4 and 0.8~$L_*$) at distances of
$360-400~\rm{kpc}$ \citep{williams10} and for the $z=0.03$ absorber,
the NASA/IPAC Extragalactic Database reports four galaxies at
$90-240$~kpc the farthest being $\gtrsim L^*_{b_j}$ and two Abell
clusters within $1.7~\rm{Mpc}$.}

Fig.~\ref{pieplot} shows the galaxy
distribution (dots) around H~2356-309 from the 2dF
Galaxy Redshift Survey \citep[2dFGRS;][]{colless}. The 2dFGRS is
redshift complete at $>90\%$ for $b_j<19$ which corresponds to $\sim
0.46~\rm{L^*_{b_j}}$ at $z\approx0.112$.  As can be seen the sightline
to the blazar passes through a $\gtrsim 30$~Mpc long filament of
galaxies connecting clusters and groups of galaxies in the same
redshift range. 
The position of the X-ray absorber ($\rm{z_{abs}}=0.1117 \pm 0.0010$: statistical and absolute wavelength scale uncertainties, added 
in quadrature) is marginally consistent (at a $\sim 1.8\sigma$ level, once both the absorber redshift uncertainty and the 
width of the galaxy distribution - FWHM = 600 km s$^{-1}$ - are properly taken into account) with the centroid of the galaxy redshift 
distribution ($\rm{z_{cen}}=0.1141$, i.e. $\sim +725~\rm{km\,s^{-1}}$ from $\rm{z_{abs}}$; see Fig. 3, bottom panel).
By conservatively assuming that all galaxies are at the same redshift of the absorber, we find that the closest galaxy is
$\sim2.2~\rm{Mpc}$ distant (the actual 3-D distance could be even larger).
This is comparable to the virial radius of a massive X-ray emitting cluster,
(e.g. \citealt{pointecouteau}). Moreover within a distance of $4$~Mpc
there are only 9 galaxies with luminosities in the range
$0.3-0.7~\rm{L^*_{b_j}}$. Hence the absorber lies in a relatively
under-dense galaxy environment. 

We checked
for the presence of possible galaxies closer to the sightline but with
unkwnown redshift. At the redshift of the absorber they would have
luminosities $0.1-0.3~L^*_{b_j}$ and be distant
$\sim200-600~\rm{kpc}$. The closest galaxies of luminosities
comparable to $\sim L^*_{b_j}$ are located at larger distances. In
particular within $1~\rm{Mpc}$ there are three objects of $0.8$, $1$ and $0.6~L^*_{b_j}$ at distances respectively of $0.6$, $0.85$, $0.9~\rm{Mpc}$. 
It is also possible that there could be a galaxy   
exactly aligned to the sightline (therefore not immediately visible) producing the absorption. 
\citet{falomo} studied the morphological properties of the host
galaxies of a sample of blazars with HST-WFPC2 including
H~2356-309. They found  a compact faint  ($m_R=22.5$) object at
$1.2^{\prime\prime}$ from the blazar position. At the redshift of the absorption 
this companion would be $\sim2.5~\rm{kpc}$ distant and have a physical
size of a fraction of kpc. We checked for its possible counterpart at
other wavelenghts. The HRC-S/LETG order 0 image does not show signs of
it. Moreover \citet{giroletti} analyzing VLA and VLBA maps do not find
any radio counterpart at these scales. 
Although there are no optical spectroscopic studies for this specific
compact object, \citet{falomo} find several of 
these ``companions'' in sub-arcsec studies of BL Lac objects statistically
hinting at a possible physical connection to the blazars themselves. 
In this context this compact companion can more likely be the
responsible of the CV line at the blazar rest-frame than of the 
intervening one.
We therefore conclude that the CV absorption that we report here, is probably produced by a genuine WHIM filament , locally free from galaxies, in a substantially 
metal-enriched intergalactic environment, and not by a nearby galaxy extended halo.

\section{Summary and conclusions}
We reported on a strong intervening CV K$\alpha$ absorber at $z=0.1117 \pm 0.0005$, serendipitously detected in the 
$\sim600~\rm ks$ Chandra HRC-S/LETG spectrum of the blazar H2356-309. The CV K$\alpha$ absorption line is detected 
at a $\sim2.9\sigma$ confidence level (accounting for the number of
independent redshift trials). 

We tentatively identify this absorber with an intervening WHIM absorber embedding a large-scale filament of galaxies extending  for $\sim 30$ Mpc, and connecting clusters and groups of galaxies at its extremes, and we rule out its possible 
association with the extended halo of a bright single galaxies at a distance lower than 2 Mpc from the absorber. 

Future FUV studies with HST-COS coupled with further deeper and redshift-complete optical spectroscopic observations of the local 
galaxy field will certainly help in shedding further light on the origin of the absorber, its metallicity and physical properties.

\begin{acknowledgments}
We would like to thank the anonymous referee for the useful comments,
Gianpiero Tagliaferri for the suggestions on the first stage of 
the manuscript and Ehud 
Behar for the wavelength calculations obtained with the HULLAC code.
This research has made use of the NASA/IPAC Extragalactic Database (NED) which is operated by the Jet Propulsion Laboratory, California Institute of Technology, under contract with the National Aeronautics and Space Administration. 
\end{acknowledgments}

\bibliography{whim}

\end{document}